\documentclass[prd,twocolumn]{revtex4}

\usepackage{graphicx, epsfig}
\usepackage{epstopdf}
\usepackage{color}
\usepackage{mathrsfs}
\usepackage{amsmath}
\usepackage{braket}
\usepackage{textcomp}
\newcommand{\be}{\begin{equation}}
\newcommand{\ee}{\end{equation}}
\newcommand{\bea}{\begin{eqnarray}}
\newcommand{\eea}{\end{eqnarray}}

\newcommand{\gapp}{\mathrel{\raise.3ex\hbox{$>$}\mkern-14mu
              \lower0.6ex\hbox{$\sim$}}}
\newcommand{\lapp}{\mathrel{\raise.3ex\hbox{$<$}\mkern-14mu
              \lower0.6ex\hbox{$\sim$}}}

\begin{document}
\title{Gravitational collapse and Hawking-like radiation of the shell in AdS spacetime}
\author{Anshul Saini, Dejan Stojkovic}
\affiliation{ HEPCOS, Department of Physics, SUNY at Buffalo, Buffalo, NY 14260-1500, USA}

 %%%%%%%%%%%%%%%%%%%%%%%%%%%%%%%%%%%%%%%%%%%%%%%%%%%%%%%

\begin{abstract}

In this paper, we study the collapse of a massive shell in 2+1 and 3+1 dimensional gravity with Anti-de Sitter asymptotics. Using the Gauss-Codazzi method, we derive gravitational equations of motion of the shell. We then use the functional Schrodinger formalism to calculate the spectrum of particles produced during the collapse. At the late time, radiation agrees very well with the standard Hawking results. In 3+1 dimensions, we reproduce the Hawking-Page transition.  We then construct the density matrix of this collapsing system and analyze the information content in the emitted radiation. We find that the off-diagonal elements of the density matrix are very important in preserving the unitarity of the system.
\end{abstract}

%%%%%%%%%%%%%%%%%%%%%%%%%%%%%%%%%%%%%%%%%%%%%%%%%%

\maketitle

\section{Introduction}

%\widetext
The existence of Hawking radiation is an outcome of quantum effects in the curved spacetime near a black hole. Although Hawking's original calculations were based on a ray-tracing method in the vicinity of the black hole horizon\cite{Hawking:1974sw, Hawking:1976ra}, the Planckian spectrum of emitted radiation has been derived using other methods as well\cite{Parikh:1999mf, Shankaranarayanan:2000qv, Gibbons:1977mu}. The thermal nature of Hawking radiation gives rise to the so-called information loss paradox and this has led to extensive research on the origin of Hawking radiation\cite{Page:1979tc, Mathur:2009hf, Lochan:2016nbs}.

In the case of the static black holes, the existence of an event horizon is crucial for particle production because the Killing vector which is timelike outside becomes spacelike inside the horizon, thus allowing for the macroscopic flux of positive energy toward infinity and negative energy inside the black hole. However, one might also consider a dynamical collapse. A time-dependent metric can also be responsible for particle production since energy is not conserved. During a gravitational collapse, the dynamical background spacetime excites the modes of the fields which propagate in that background, thereby producing radiation\cite{Vachaspati:2006ki,Dabrowski:2016tsx}. This particle production, also known as pre - Hawking radiation, should approach Hawking radiation as the collapsing object reaches its own event horizon. Hence, it is crucial to understand particle production due to a collapsing object in the vicinity of its own horizon. Additionally, a static event horizon does not form within finite time with respect to an outside observer, and hence understanding pre - Hawking radiation is very important.

Black hole physics in Anti-de Sitter (AdS) spacetimes has attracted a lot of attention, partially because of the non-trivial structure of the curved AdS space, and partially due to the possible AdS/CFT correspondence. Most of the existing work has focused on pre-existing black holes in AdS space. Although there is some work on the gravitational collapse leading to the formation of black holes, the bulk of it is on the collapse of a scalar field in the curved background (see e.g. \cite{Pretorius:2000yu,Allahyari:2014lta}). In addition, there is virtually no work on the quantum radiation, which accompanies the gravitational collapse of an object.

In this paper, we study gravitational collapse of a dust shell in the presence of a test scalar field in an asymptotically AdS spacetime. In the commonly used Bougliubov method, knowledge of the explicit form of the modes is critical in order to calculate the mismatch of modes at early and late times of the collapse. However, finding the explicit form of modes is a challenging task. To circumvent this problem, we applied the functional Schrodinger formalism to study the emitted particle spectrum \cite{Vachaspati:2007hr,Greenwood:2010mr,Greenwood:2009gp,Vachaspati:2007ur,Kolopanis:2013sty}. This method is suitable for time-dependent backgrounds since the time-dependent wavefunctional contains information about all field excitations and their time evolution. In addition, this method does not require knowledge of the explicit decomposition of the field into the particle modes. We find that the radiation excited during the collapse approaches the Planckian spectrum and matches the standard Hawking's result, as the collapsing shell approaches its own horizon radius. This method also allowed us to construct the total density matrix of the excited modes in a way similar to the Schwarzschild black holes\cite{Saini:2015dea}.

This paper is organized as follows: In section II, we study the classical collapse of the dust shell in 2+1 dimensional gravitational spacetime using the Gauss-Codazzi method. This space-time admits the existence of an event horizon, giving rise to black hole solutions known as the BTZ black holes. In section III, we apply the functional Schrodinger formalism and derive the particle spectrum and temperature of the emitted radiation during late times. In section IV, we construct the density matrix of emitted quanta to analyze its information content. In section V, we repeat the analysis for 3+1 dimensional asymptotic AdS spacetime. Finally, in section VI, we conclude our findings.

\section{ 2+1 AdS spacetime: dynamics of the classical shell collapse}
\label{sc}

In this section, we examine the classical collapse of a dust shell in 2+1 dimensional gravity spacetime. The primary objective is to derive an equation of motion governing the dynamics of the collapsing shell. This equation of motion is essential to obtain the radiation spectrum. In \cite{Banados:1992wn}, it was shown that a $2+1$ dimensional asymptotically AdS spacetime accommodates a spherically symmetric solution of Einstein equations. The  metric outside the collapsing shell can be written as

\be
ds^2 = - \left(\frac{r^2}{l^2} -  M \right) dt^2 + \frac{1}{ \left(\frac{r^2}{l^2} -  M \right)} dr^2 + r^2 d\phi^2
\ee

where $M$ is the mass of the shell and $l$ is a parameter related to the cosmological constant $\Lambda$ as $\Lambda = -1/l^2$. It is evident from the metric that the event horizon is located at $r = M l^2 $. This  vacuum solution of the Einstein equation is commonly known as the BTZ black hole.  BTZ black holes differs fundamentally  from the  Schwarzschild black holes due to the absence of the singularity at the center\cite{Carlip:1995qv}. BTZ black holes have positive specific heat, and as a result, they can exist in stable equilibrium with the environment. This metric becomes an usual AdS spacetime for $M = -1$. The metric inside the collapsing shell is obtained by putting $M = 0$, which is also known as a "massless black hole". Thus, the form of metric inside the shell is given by
\be
ds^2 = - \left(\frac{r^2}{l^2}\right) dT^2 + {\left(\frac{r^2}{l^2} \right)}^{-1}dr^2 + r^2 d\phi^2 .
\label{inmetric}
\ee
The metric on the shell is
\be
ds^2 = - d \tau^2 + r^2 d\phi^2 .
\label{onshellmetric}
\ee
Since the metric must be continuous, we can match the outside and inside metric at the shell, to get the relationship between the time coordinates as
\be
\frac{dT}{d\tau} = \frac{1}{B_{in}}\sqrt{ B_{in}+ \left(\frac{dR}{d\tau}\right)^2} ,
\label{Ttau}
\ee
and
\be
\frac{dt}{d\tau} = \frac{1}{B_{out}}\sqrt{ B_{out}+ \left(\frac{dR}{d\tau}\right)^2} ,
\label{ttau}
\ee
where $B_{in} =(\frac{R^2}{l^2})$ and $B_{out} = (\frac{R^2}{l^2} -  M )$.  To determine the equation of motion of the shell, we apply Gauss-Codazzi method for the surface layers\cite{Ipser:1983db, Schmidt}. As the shell is moving, it curves the spacetime and creates discontinuity in the extrinsic curvature ($K$).  Einstein equations for this case in their general form become
\be
\left[ K^i_j \right]  - \delta_{j}^{i} Tr[K] = 8 \pi  S^i_j ,
\ee
where
\be
S^{\alpha}_{\beta} =  \int_{\epsilon}^{\epsilon} T^{\alpha}_{\beta} dn
\ee
 is the surface stress energy tensor, and [K] is the discontinuity in the extrinsic curvature. We can take the trace of both sides of the equation and obtain an expression for $[K^i_j]$ as
\be
\left[ K^i_j \right] = 8 \pi \left( S^i_j - \frac{1}{2} \delta^i_j S^k_k \right) .
\ee
Now, we can substitute the expressions for our concrete case. The surface energy tensor for the collapsing shell is $S^{\alpha \beta} = \sigma u^{\alpha} u^{\beta} $, where $\sigma$ is the surface energy density of the shell. Substituting this in the above equation gives
\be
\left[ K^i_j \right] = 8 \pi \sigma \left( u_i u_j -  {}^{(2)}g_{ij} \right) .
\label{extrincur}
\ee
Due to the absence of stress energy tensor outside the shell, the equation of motion becomes
\be
\frac{d \sigma}{d \tau} + \sigma u^i \mid_i =0 ,
\ee
where $u^i \mid_i$ is a covariant derivative in the given curved background (in our case AdS). After simplification, it yields $(\sigma R )_{,\tau} = 0$, which  upon integration over the whole space gives $2\pi \sigma R = \mu$. Here, $\mu$ is a constant of motion which can be interpreted as the rest mass energy of the shell. Using Eq. (\ref{extrincur}), the expression for $[K_{\phi \phi}]$ can be calculated as
\bea
[K_{\phi \phi}] &=& 8 \pi \sigma (u_\phi u_\phi + {}^{(2)}g_{\phi \phi} )\\
                      &=& 4 \mu r
\label{extrincur1}
\eea
This equation allows us to calculate  $[K_{\phi \phi}]$  using the energy associated with the shell. Alternatively, one can figure out the expression of $[K_{\phi\phi}] $  through the spacetime geometry. Extrinsic curvature is defined as $ K_{\phi \phi} = - n_{\phi; \phi} = - r n^r $. Plugging this in the left hand side of the Eq. (\ref{extrincur1}), gives us
\be
n^{r+} - n^{r-} = - 4 \mu .
\label{normaleq}
\ee
To compute the normal vector components, we use the identities $n.n = - u.u = 1$ and $u.n =0$. Consider $n^{r+}$, which is the unit normal vector pointing outside the shell in $r$-direction.  For the metric outside the shell, these identities lead to three equations
\bea
\left(\frac{r^2}{l^2} -  M \right)(u^t)^2 - \frac{1}{\left(\frac{r^2}{l^2} -  M \right)}(u^r)^2 = 1\\
-\frac{1}{\left(\frac{r^2}{l^2} -  M \right)} (n_t)^2 + \left(\frac{r^2}{l^2} -  M \right)(n_r)^2 = 1\\
u^t n_t + u^r n_r = 0  .
\eea
Eliminating  $n_t$ and $u^t$, gives $n_{r}^{+}$ as
\be
n_r^{+} = \sqrt{\frac{1 + \frac{(u^r)^2}{\left(\frac{r^2}{l^2} -  M \right)}}{\left(\frac{r^2}{l^2} -  M \right)}} .
\ee
Near the shell, $u^{r} = \frac{dR}{d\tau}$. Substituting this into the previous equation gives the contravariant component of the normal vector
\be
n^{r+} = \sqrt{\dot{R}^2 +\frac{R^2}{l^2} -  M} .
\ee
Similarly, $n^{r-}$ (which points inside) can be calculated by following the same procedure and setting $ M = 0 $ as
\be
n^{r-} = \sqrt{\dot{R}^2 +\frac{R^2}{l^2}} .
\ee
Plugging $n^{r+}$ and $n^{r-}$ into the Eq. (\ref{normaleq}) gives us the equation of motion of the collapsing shell as
\be
\frac{dR}{d \tau} = \sqrt{\left(\frac{ M+ 16 \mu^2 }{8 \mu}\right)^2 -\frac{R^2}{l^2}}
\ee
This equation is written in terms of the shell time coordinate $\tau$, which is the proper time coordinate of an observer moving together with the shell. However, we are interested in the collapse from the outside observer's perspective. Using Eq. (\ref{ttau}), we can transform the time coordinate from $\tau$ to $t$. The dynamics of the shell described by an outside observer is
\be
\frac{dR}{dt} = B \left( 1 +B  \left(\frac{dR}{d \tau}\right)^{-2}\right)^{-\frac{1}{2}} .
\ee
Radiation emitted from the shell during its gravitational collapse will depend on this dynamical equation. We will do this in the next section.

\section{2+1 dimensional AdS spacetime: radiation from the collapsing shell}

\begin{figure}[htpb]
\begin{center}
  \includegraphics[height=0.28\textwidth,angle=0]{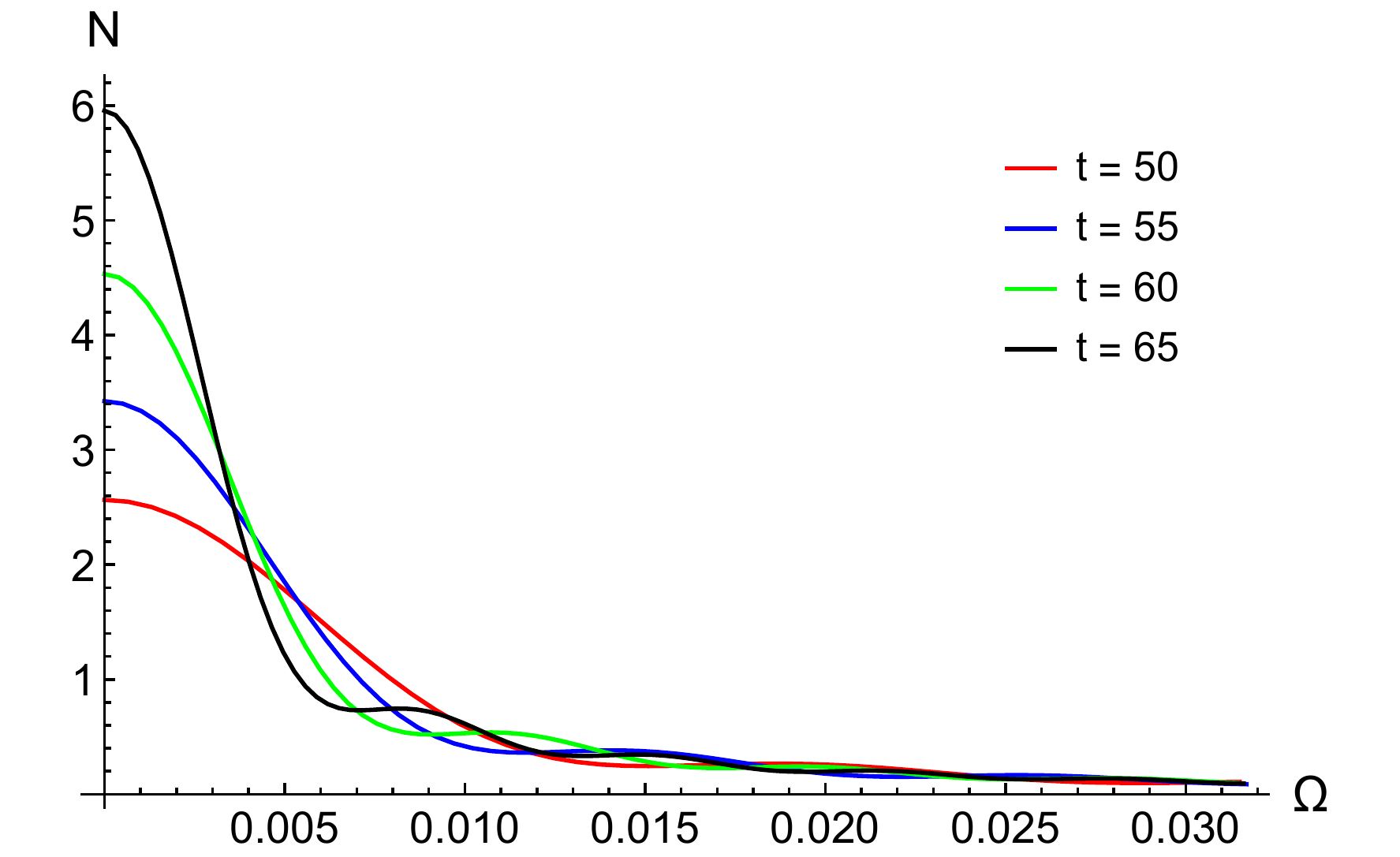}
\caption{Occupation number of the excitations given by Eq. (\ref{on}) is plotted as a function of frequency at different times. Clearly, as time increases, distribution becomes more and more Planckian. }
\label{Nvsomega}
\end{center}
\end{figure}

\begin{figure}[htpb]
\begin{center}
  \includegraphics[height=0.28\textwidth,angle=0]{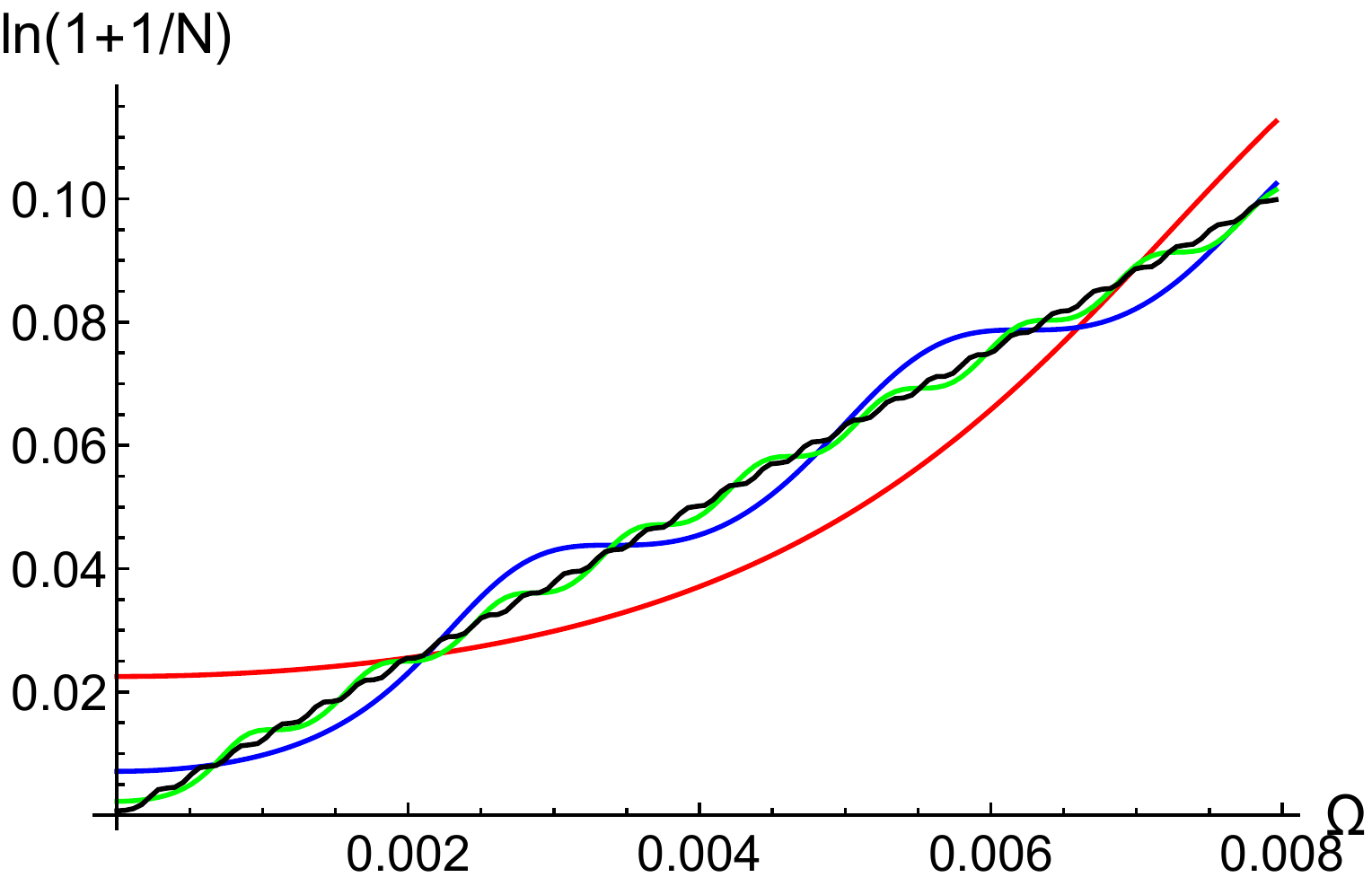}
\caption{In this figure, we plot $\ln(1 + 1/N)$ as a function of $\Omega$ at different times. The spectrum is not thermal at early times due to large fluctuations. However, as time progresses it becomes more and more thermal. At late times, the slope of the plot is inverse temperature. We also calculate the ratio of the temperature obtained by fitting the curve and the standard Hawking temperature. We see that the ratio approaches unity as time progresses.}
\label{lnNvsomega}
\end{center}
\end{figure}

We consider a minimally coupled scalar field in the collapsing  shell background. As mentioned earlier, the time dependent geometry should lead to creation of particles.  The scalar field modes placed in the time-dependent geometry get excited out of vacuum. Since, strictly speaking, particles are well defined only in static spacetimes where one can define a proper timelike Killing vector, we refer to these excitations as modes rather than particles in the usual sense. The action for a minimally coupled scalar field propagating in a curved spacetime is
\be
S =  \int{ d ^{3} x \sqrt{-g} \frac{1}{2} g^{\mu \nu } {\partial}_{\mu} \phi  {\partial}_{\nu} \phi} .
\ee
Inside the shell, the metric is given by Eq. (\ref{inmetric}). The corresponding action of the scalar field is
\be
S_{in} =  \pi \int{dT} \int_{0}^{R(t)} dr r \left[ -  \frac{({\partial}_{T} \phi)^{2}}{ r^2/l^2} +\frac{r^2}{l^2} ({\partial}_{r} \phi)^{2} \right] .
\ee
Since, we are interested in the spectrum of the excited modes with respect to an outside observer,  $S_{in}$  is re-written in terms of outside time coordinate ($t$) as
\be
S_{in} = \pi \int{dt} \int_{0}^{R(t)} dr r \left[ -  \frac{({\partial}_{t} \phi)^{2}}{(r^2/l^2) \dot{T}} + (\frac{r^2}{l^2})\dot{T}({\partial}_{r} \phi)^{2} \right] ,
\ee
where $\dot{T} = \frac{dT}{dt}$. The expression for $\dot{T}$ can be evaluated from Eq. (\ref{Ttau}) and Eq. (\ref{ttau})
\be
\frac{dT}{dt} = \frac{B_{out}}{B_{in}}\sqrt{\frac{B_{in} + R_{\tau}^2}{B_{out} + R_{\tau}^2}}  .
\ee
Plugging the relations for $B_{in}$, $B_{out}$ and $R_{\tau}$ yields
\be
\lambda(t) = \frac{dT (t)}{dt} = \frac{(16 \mu^2 + M)}{(16 \mu^2 - M)}\left( 1 - \frac{M l^2}{R(t)^2}\right) .
\label{lambda}
\ee
Similarly, the action outside the shell is
\be
S_{out} =  \pi \int{dt} \int_{R(t)}^{\infty} dr r \left[ -  \frac{({\partial}_{t} \phi)^{2}}{(\frac{r^2}{l^2} -  M)} +\left(\frac{r^2}{l^2} -  M\right) ({\partial}_{r} \phi)^{2} \right].
\ee
The total action can be obtained by summing up both contributions. Since $\dot{T}$ approaches zero faster than $\arctan (R^2 - Ml^2)$, we can ignore the kinetic term in $S_{out}$ and  the gradient term in  $S_{in}$. The final expression becomes
\bea
S_{tot} \sim && \pi \int dt \left( - \frac{1}{\lambda} \int_{0}^{R_s} dr r  \frac{1}{\left(r^2/l^2\right)} ({\partial}_{t} \phi)^{2} + \right. \\
&& \left. \int_{R_s}^{\infty} dr r \left( \frac{r^2}{l^2} -  M\right) ({\partial}_{r} \phi)^{2}\right) \nonumber  .
\eea
We can now decompose the scalar field in terms of the spherically symmetric basis functions as
\be
\phi  = \sum_{k} a_{k} (t) f_{k} (r) .
\ee
The time dependence of the scalar field  is absorbed in the coefficients $a_k (t)$, while $f_k (r)$ is a set of orthonormal basis functions. The modes $a_k (t)$ are the dynamical variables in our context, and the explicit form of $f_k (r)$ is not needed in our approach. In terms of these functions, the action takes the form
\be \label{action}
S = \int dt \left(  \frac{1}{2\lambda} \frac{d{a}_{k}}{dt} A_{kk\rq{}}\frac{d{a}_{k\rq{}}}{dt}  - \frac{1}{2}  a_{k} B_{kk\rq{}}a_{k\rq{}} \right) .
\ee
The matrices $A_{kk'}$ and $B_{kk'}$ are given by
\bea
&& A_{kk\rq{}} = - 2 \pi  \int_{0}^{R_s} dr r \frac{1}{\left(r^2/l^2\right)} f_k (r) f_{k\rq{}} (r)\\
&& B_{kk\rq{}} = - 2 \pi \int_{R_s}^{\infty} dr r \left( \frac{r^2}{l^2} -  M \right) f\rq{}_k (r) f\rq{}_{k\rq{}} (r) .
\eea
Once the action is known, we can proceed with the usual quantization procedure. From the action in Eq.~(\ref{action}), we can construct the Hamiltonian of the modes, and then write the Schrodinger equation for the wavefunctional $\psi(a_k, t)$ as
\be
\left[ \lambda \frac{1}{2} \Pi _k  {(A^{-1})}_{kk\rq{}} \Pi_{k\rq{}}  + \frac{1}{2} a_k B_{kk\rq{}} a_{k\rq{}} \right] \psi(a_k , t) = i \frac{\partial \psi(a_k, t)}{\partial t} .
\ee
Here $\Pi$ is the generalized momentum corresponding  to the dynamical variable $a_k (t)$
\be
\Pi_k = - i \frac{\partial}{\partial a_k} .
\ee
Since matrices $A$ and $B$ are symmetric and real, both can be diagonalized simultaneously by applying the principle axis transformation. Then, one can write the Schrodinger equation for the eigen mode
\be
\left[ -
\lambda \frac{1}{2\alpha} \frac{{\partial}^2}{\partial b^2} + \frac{1}{2} \beta  b^2 \right] \psi(b, t) = i \frac{\partial \psi (b,t)}{\partial t} .
\label{schrodinger}
\ee
where $b$ is the eigenmode (a linear combination of the original modes $a_k$), while $\alpha$ and $\beta$ are eigenvalues of the matrices $A$ and $B$ respectively. Simultaneous diagonalization of matrices $A$ and $B$ immensely simplify the problem since all the eigenmodes are decoupled now. In order to solve this partial differential equation, let us perform a coordinate transformation to a new time variable $\eta$ as
\be
 \label{eta}
\eta = \int_{0}^{t} dt \lambda
\ee
where $\lambda$ is given by Eq. (\ref{lambda}). In terms of $\eta$,  the Schrodinger equation can be written as
\be
\left[ - \frac{1}{2\alpha} \frac{{\partial}^2}{\partial b^2} + \frac{\alpha}{2} {\omega}^2 (\eta) b^2 \right] \psi(b, \eta) = i \frac{\partial \psi (b,\eta)}{\partial \eta} ,
\ee
where
\be\label{baromega}
{\omega}^2(\eta) =\left(\frac{\beta}{\alpha}\right) \frac{1}{\lambda} \equiv \frac{{\omega_0}^2}{\lambda} .
\ee
This is an equation of a harmonic oscillator, hence the system is reduced to a set of decoupled harmonic oscillators with time dependent frequency. The above equation admits an exact solution in the form of
\be
\psi (b, \eta) = e^{i \delta (\eta)}{ \left[\frac{\alpha}{\pi  \theta^2}\right]}^{\frac{1}{4}} \exp \left[  i \left( \frac{\theta_\eta}{\theta} +\frac{i}{\theta^2}\right) \frac{\alpha b^2}{2} \right]
\ee
where  $\theta$ is the solution of a differential equation
\be
\theta_{\eta \eta} + {\omega}^2 (\eta)  \theta = \frac{1}{\theta^3}
\ee
with initial conditions
\bea
\theta(0) = \frac{1}{\sqrt{\omega_0}},  \ \ \  \theta_\eta (0) = 0 .
\eea
Different choice of initial conditions correspond to different modes $b$ (though all the modes $b$ satisfy the same differential equation).
Hence the partial differential equation has been reduced to an ordinary differential equation which can be solved numerically. To calculate the spectrum of the excited  modes, the wavefunctional is expanded in terms of the harmonic oscillator basis functions  $\zeta_n (b)$ as
\be
\psi (b,t) =  \sum_{n} c_n (t) \zeta_n (b) ,
\ee
where the coefficients $c_n$ are given by
\be
c_n(t) =  \int db \ {\zeta_n}^{*}(b) \psi (b,t) .
\label{coeffdef}
\ee
While the choice of this basis appears natural for our case, this set is infinite which makes it inconvenient for numerical calculations. However, as we will find soon, higher excited states (corresponding to higher values of the index $n$) are increasingly suppressed so they are not crucial for establishing our conclusions as long as a large enough value of $n$ is taken into account. Integrating Eq.(\ref{coeffdef}), the structural form of the coefficients $c_n(t)$ is calculated in terms of time and frequency as
\begin{equation}
  c_n(t)=\frac{(-1)^{n/2}e^{i\alpha}}{(\Omega \lambda^{-1} \theta^2)^{1/4}}\sqrt{\frac{2}{P}}\left(1-\frac{2}{P}\right)^{n/2}\frac{(n-1)!!}{\sqrt{n!}}.
\end{equation}
where  $P$ is
\begin{equation}
  P=1-\frac{i\lambda}{\Omega}\left(\frac{\theta_\eta}{\theta}+\frac{i}{\theta^2}\right).
\end{equation}
The occupation number of the excited modes is
\be \label{on}
  N(t,\Omega)=\frac{\Omega\theta^2}{4 \lambda}
  \left[\left(1-\frac{\lambda}{\Omega\theta^2}\right)^2+\left(\frac{\theta_{t}}{\Omega\theta}\right)^2\right] .
\ee
It is instructive to compare this occupation number with the standard Planckian distribution
\be
N(\omega) = \frac{1}{e^{\beta \omega} -1}
\label{planckspect}
\ee
where $\beta = 1/T $. For the pre-existing BTZ black hole, the Hawking temperature is  $T_{H} =\sqrt{M}/{2 \pi l}$.
In Fig.~\ref{Nvsomega}, we plot the occupation number from Eq.(\ref{on}) as a function of frequency at different times. It is clear from the plot that the distribution approaches a Planckian one, as time progresses.
From Eq.~(\ref{planckspect}), we see  that $\ln(1 + 1/N) = \beta \omega$, so if we plot $\ln(1 + 1/N)$ vs. $\omega$, the slope of the graph will be the inverse temperature. In Fig.~\ref{lnNvsomega},  the ln(1 + 1/N) is plotted as a function of $\Omega$ to estimate the temperature of the emitted radiation with respect to an asymptotic observer. At early times, the fluctuations are too large to define temperature, which means that the spectrum is not Planckian. This is fine since we expect the Planckian spectrum only when the collapsing object is very close to its own (future) horizon. Fluctuations die out with time, and at late times temperature can be extracted from the slope. In the fitted frequency range, the temperature of the emitted radiation matches exactly the Hawking temperature. The fitted $\Omega$ range lies from $0$ to $0.1 T_{H}$ which is small in the outside observer reference frame, but the corresponding initial frequency of the mode, $\omega_0$, at the time of mode creation is large (due to the redshift $\Omega = \omega_0 \sqrt{\lambda}$). As the shell approaches its own horizon,  $\lambda \rightarrow 0$, so the redshift is huge at late times. We note that in the original Hawking calculations, geometrical optics approximation was used, which is valid only at very high frequencies. Therefore, it is natural to expect a perfect agreement in that regime.

However, the advantage of our method is that we keep track of all the frequencies.  For very low frequencies ($\Omega << T_{H}$), the spectrum may not look Plankian at first, but fluctuations become increasingly smaller as the shell approaches its own horizon, so at $t\rightarrow \infty$, it becomes Planckian for all $\Omega$ lower than $\approx 0.1 T_{H}$. Unfortunately, our analysis breaks down at very large values of $\Omega$ because the suppression is not exponential as expected in a Planckian spectrum. This might be due to the vacuum polarisation effect, and suitable regularization should remove it.  \\

\begin{figure}[htpb]
\begin{center}
  \includegraphics[height=0.28\textwidth,angle=0]{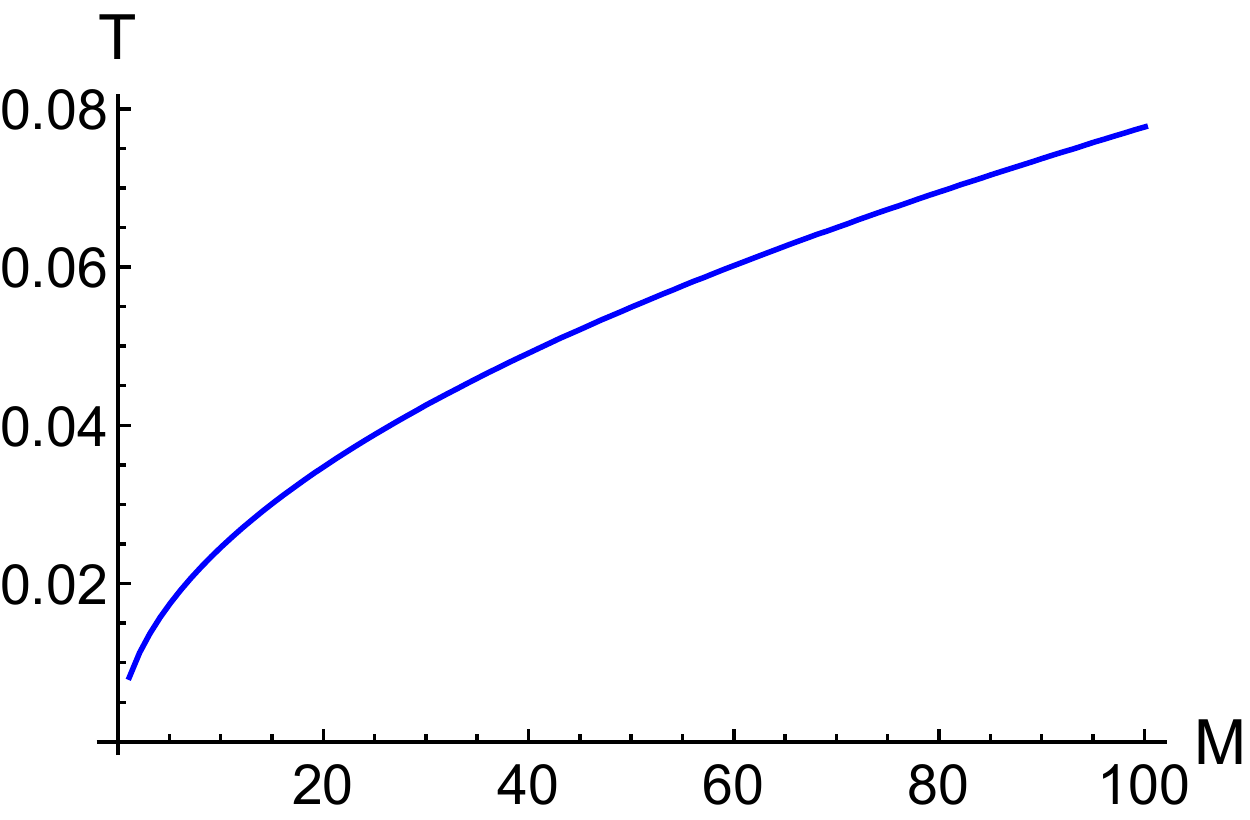}
\caption{We plot the temperature of emitted radiation as a function of the mass of the collapsing shell, for the fixed AdS parameter $l= 20$, at late times. The dependence $T \propto \sqrt{M}$ is in agreement with the standard results. }
\label{temvsmass}
\end{center}
\end{figure}

\begin{figure}[htpb]
\begin{center}
  \includegraphics[height=0.28\textwidth,angle=0]{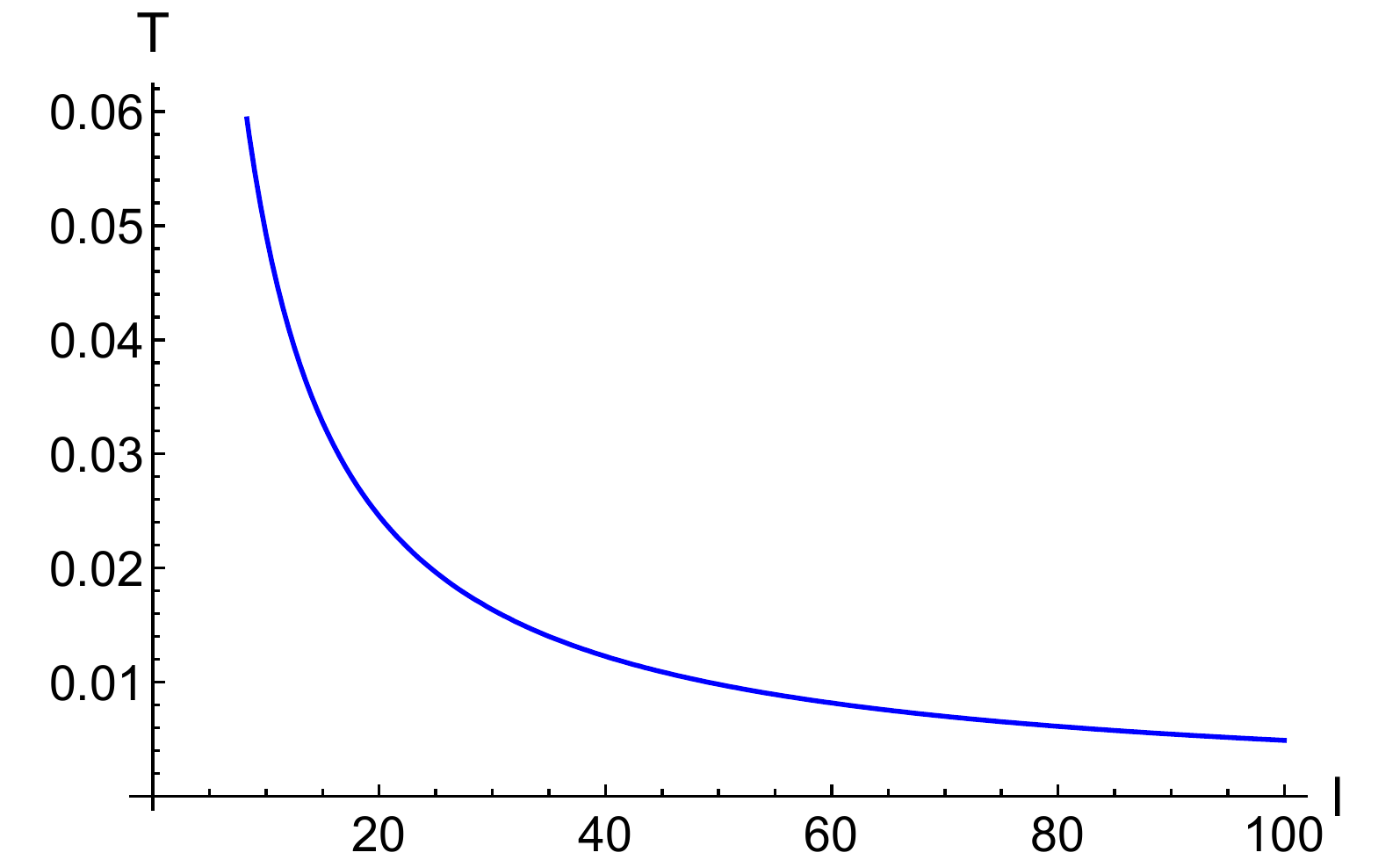}
\caption{We plot the temperature of emitted radiation as a function of the AdS parameter, $l$, keeping the fixed mass, $M =10$. The temperature is inversely proportional to $l$, thus matching the standard results.}
\label{temvsadspara}
\end{center}
\end{figure}

As mentioned before, the temperature of  BTZ black holes follow $T \propto \sqrt{M}/l$. In Fig.~\ref{temvsmass}, we plot the temperature of the emitted radiation as a function of the mass of the collapsing shell at late times. We get a very good agreement with the known result for the BTZ black holes where $T \propto \sqrt{M}/l$. Similarly, in Fig.~\ref{temvsadspara}, we plot the temperature of the collapsing shell as a function of the AdS parameter $l$. We find that the temperature is inversely proportional to $l$, thus matching standard results again.

At the end of this section, we note that we calculated only the occupation number of the excited modes. We did not calculate their further evolution as they try to penetrate the potential barrier  in order to escape to infinity (i.e. greybody factors), nor we calculated what happens if the modes at infinity encounter some non-trivial boundary conditions (e.g. the reflective boundary conditions, which would, in turn, reflect all the modes back to the collapsing object).

\section{Density matrix and information content}

\begin{figure}[htpb]
\begin{center}
  \includegraphics[height=0.28\textwidth,angle=0]{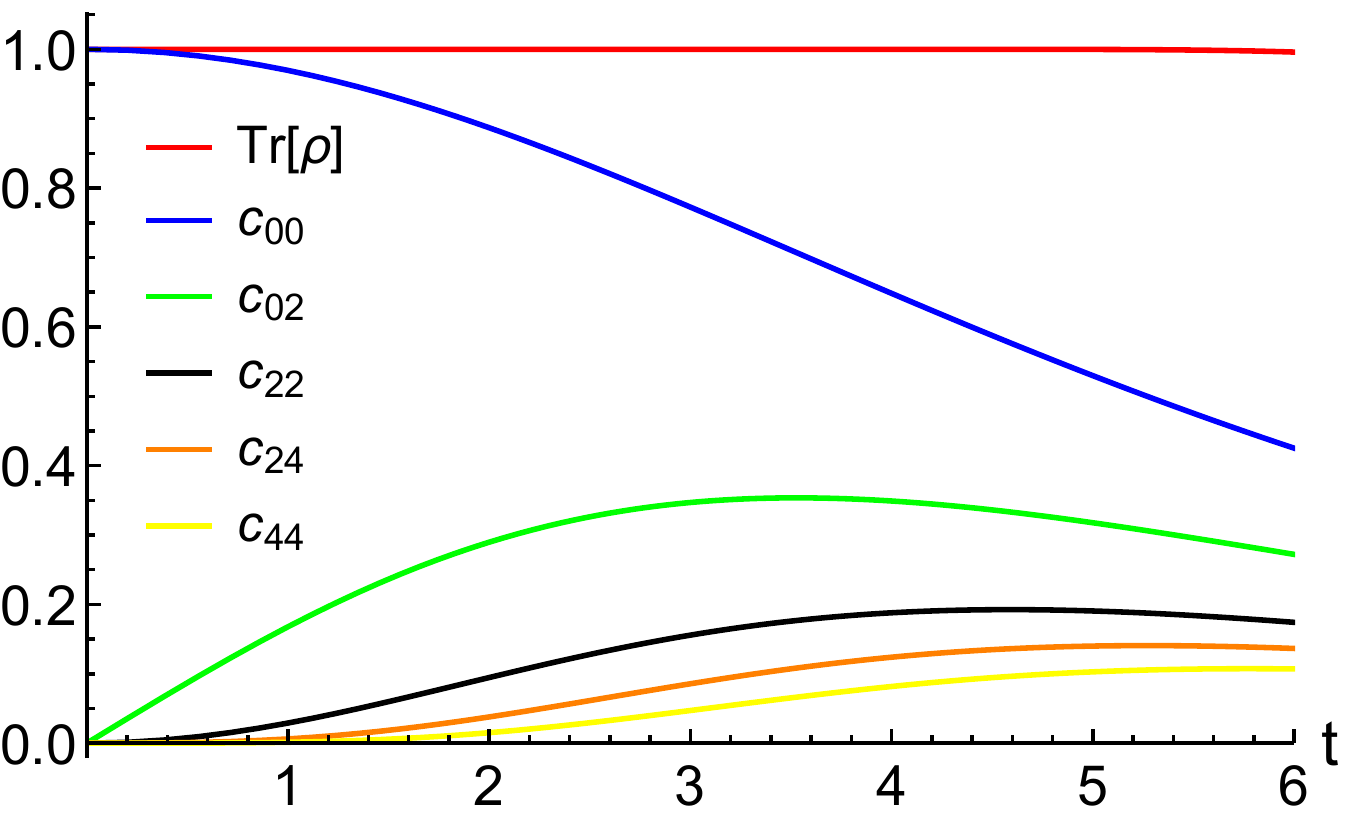}
\caption{We plot a first few non vanishing elements of the density matrix as a function of time at a fixed frequency ($\Omega = 0.0001$) for $M =100$ and $l =20$. Initially, $c_{00}$ is unity indicating that the system starts from vacuum. As time progress, $c_{00}$ decreases and excited states increase implying that the modes are excited due to the time dependent geometry. It should be noted that the off diagonal terms are of the same order as the diagonal terms hence correlations between the modes are non negligible. }
\label{2dimelementsvstime}
\end{center}
\end{figure}

\begin{figure}[htpb]
\begin{center}
  \includegraphics[height=0.28\textwidth,angle=0]{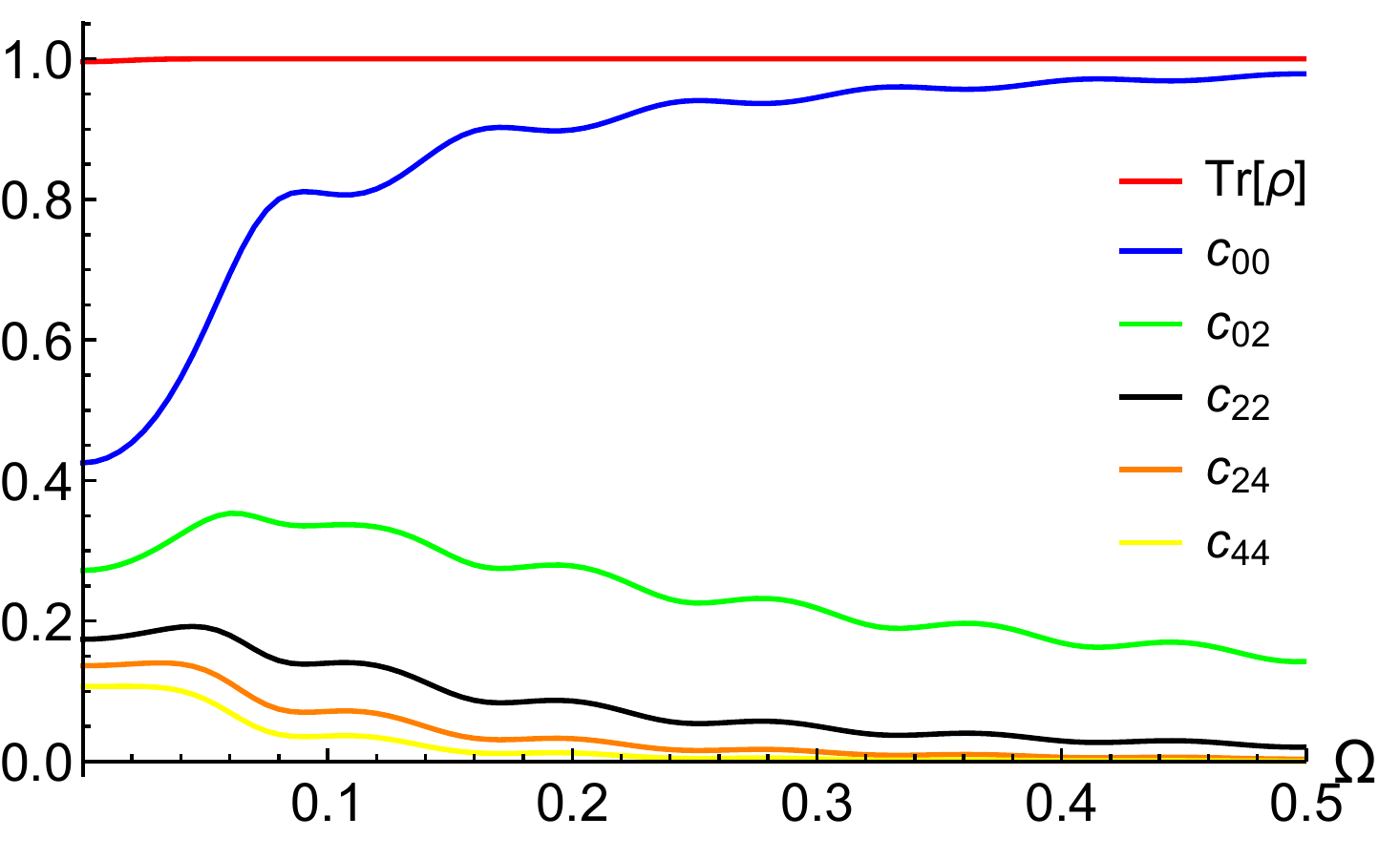}
\caption{We plot the density matrix elements as a function of frequency at a fixed time ($t =6$) for $M =100$ and $l =20$. We see that as frequency increases $c_{00}$ become dominant, indicating that the higher states are difficult to excite at high frequencies. }
\label{2dimelementsvsomega}
\end{center}
\end{figure}

Although the spectrum of the excited modes approaches Planckian distribution as the shell approaches its horizon, this does not tell us anything about the information content of the emitted radiation.  Occupation number has a diagonal form and does not contain information about the correlations between the emitted modes. In this section, we study the full density matrix, including the off-diagonal elements.  The density matrix is defined as
\be
\hat{\rho} =  \ket{\psi}\bra{\psi} ,
\ee
which in a harmonic oscillator basis takes the form
\be
\hat{\rho}  = \sum_{m,n} c_{mn}\ket{ \phi_m}\bra{ \phi_n} ,
\ee
where $c_{mn}$ are the time-dependent coefficients containing the complete information about the time evolution of the system. The diagonal elements in the density matrix give the mode occupation number, and they are always real. Off-diagonal elements give us the correlations among the modes and are generally complex. Since $H_n(0) = 0$ for odd values of $n$, all odd terms vanish, hence we have this form
\[
\rho=
  \begin{bmatrix}
    c_{00} & 0 & c_{02} & 0 & .. \\
      0  &  0&0&0&..\\
    c_{20} & 0 & c_{22} & 0 & ..\\
    0 & 0&0&0&..\\
    . & ..&..&..&..\\
    . & ..&..&..&..\\
  \end{bmatrix}
\]
In Fig.~\ref{2dimelementsvstime},  we plot a first few non vanishing elements of the density matrix as a function of time at a fixed frequency ($\Omega =0.0001$) for $M =100$, $\mu =500$ and $l=20$. As expected, the trace of all diagonal elements is unity. Initially, $c_{00}$ is unity and all other excited states are zero indicating that the system is in the vacuum state. As time progresses, higher states increase, while $c_{00}$ decreases which means that the time dependent background is exciting the field modes. Importantly, off diagonal terms are of the same order as diagonal terms, and can not be ignored. In Fig.~\ref{2dimelementsvsomega}, various elements of the density matrix are plotted as a function of frequency at a fixed time($t =6$). We found that at low frequencies the higher excited states are of the same order of $c_{00}$, and as frequency increases $c_{00}$ becomes more and more dominant with respect to the higher excited states. This is easy to explain because high frequency modes are more difficult to excite.

\begin{figure}[htpb]
\begin{center}
  \includegraphics[height=0.28\textwidth,angle=0]{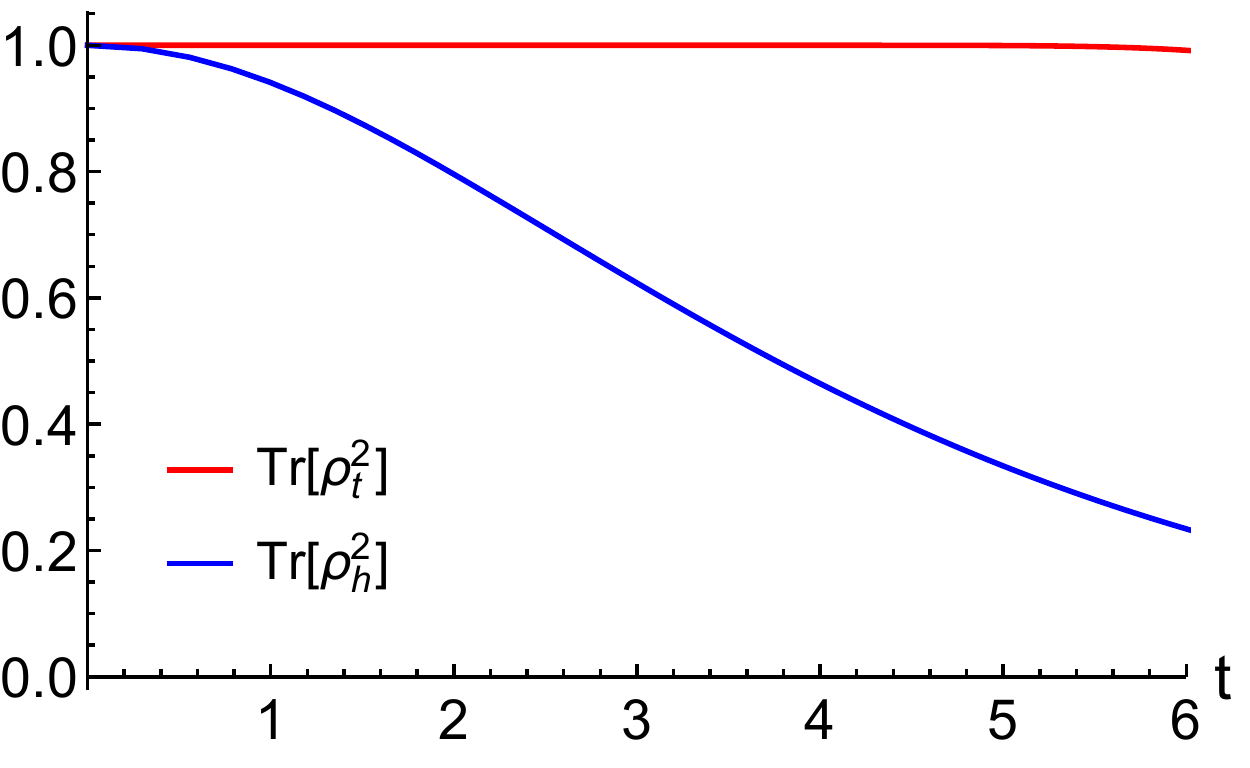}
\caption{Tr[$\rho^2$] is plotted with time at $\Omega = 0.0001$. Tr[$\rho^2_h$] contains only the diagonal elements of the matrix whereas Tr[$\rho^2_t$] contains all the elements. Tr[$\rho^2_{h}$] decreases with time suggesting that the system is going from pure to thermal/mixed state. However, Tr[$\rho^2_t$] remain unity at all time which means correlations among particles are significant enough to preserve unitarity of the system. }
\label{2dimrhovstime}
\end{center}
\end{figure}
\begin{figure}[htpb]
\begin{center}
  \includegraphics[height=0.28\textwidth,angle=0]{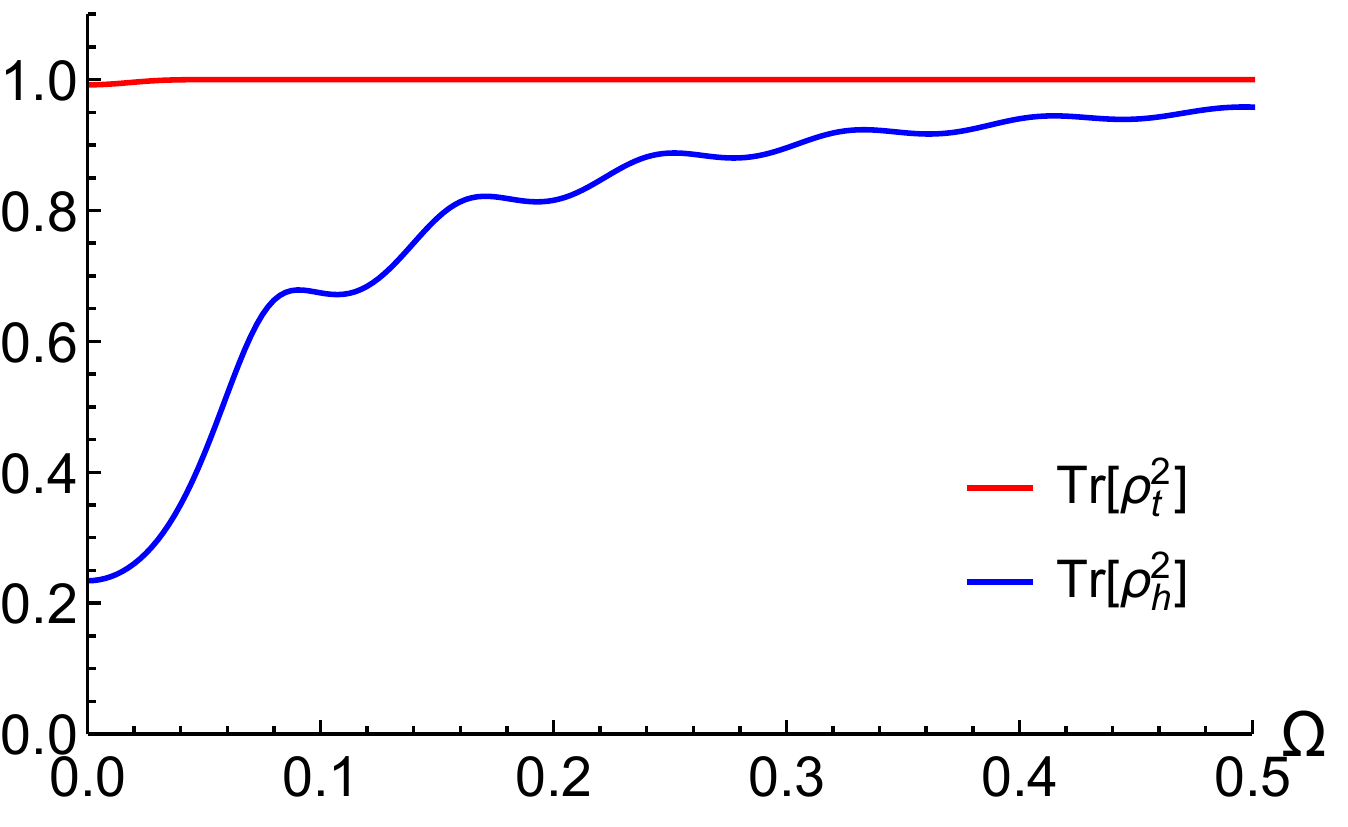}
\caption{Here we plot Tr[$\rho^2_h$] and Tr[$\rho^2_t$] as a function of omega at $t = 6$. As frequency decreases Tr[$\rho^2_h$] decreases indicating that emitted radiation is thermal at low frequencies. However, Tr[$\rho^2_t$] remains unity at all frequencies.}
\label{2dimrhovsfreq}
\end{center}
\end{figure}

Since we are interested in information evolution of the system, Tr[$\rho^2$] is plotted as a function of time at the fixed frequency. If Tr[$\rho^2$] is equal to one, then the system is in a pure quantum state. On the other hand, if it vanishes then the system is completely thermal (mixed state) and contains no information about the initial state.  Since the off-diagonal elements are non-negligible, for comparison we calculated Tr[$\rho^2_h$]  containing only the diagonal elements and  Tr[$\rho^2_t$] containing all the elements. In Fig.~\ref{2dimrhovstime}, we plot  Tr[$\rho^2_t$] and  Tr[$\rho^2_h$] as a function of time at $\Omega =0.0001$. Tr[$\rho^2_h$] decreases with time suggesting that the system is evolving from a pure to a mixed state. However, Tr[$\rho^2_t$] remains unity the whole time which means that unitarity is preserved if the off-diagonal elements taken into account. In Fig.~\ref{2dimrhovsfreq}, we plotted Tr[$\rho^2$] as a function of frequency at $t =6$. We find that Tr[$\rho^2_h$] is unity at high frequencies and decreases as frequency goes down. This is expected because at high frequencies the system is dominated by ($c_{00}$). However, Tr[$\rho^2_t$] remains unity regardless of the frequency.

We comment here on implications of this result to the information loss paradox. We demonstrated that Tr[$\rho^2_t$] remains unity during the whole evolution.
However, to claim that there is no information loss, it is crucial for an observer to be able to see the whole
density matrix, i.e. that he is able to measure all the modes and their correlations at least in principle. In the standard Schwarzschild case, when calculating radiation from a pre-existing horizon, one has to trace out the infalling modes which eventually get lost in the singularity. This leaves us with an incomplete density matrix, and unitarity is lost. Black holes in 2+1 dimensional AdS spacetime are special in that sense because they do not have a singularity at the center. The modes are therefore not irretrievably lost at the singularity, though they might be hidden beyond the horizon for some time. Since they are not destroyed in the singularity, but presumably only trapped for some finite time, they will be re-emitted when the original black hole losses enough of its mass. This would then imply that  2+1 dimensional AdS black holes may not suffer from the information loss problem. However, to verify that this indeed happens, one would have to solve the full problem including the backreaction from the radiation to the collapsing object.

\section{3+1 dimensional AdS spacetime: dynamics of the classical shell collapse and quantum radiation}
In this section, we consider the collapse of the massive shell in an asymptotic AdS space in 3+1 dimensions. The primary motive to repeat the calculations is the fact that the properties of black holes in 3+1 dimensional AdS spacetime are significantly different from the 2+1 dimensional BTZ black holes. The metric outside the collapsing shell in 3+1 dimensions is
\be
ds^2 = - \left( 1 + \frac{r^2}{l^2} - \frac{2  M}{r} \right) dT^2 + \frac{1}{ \left( 1 + \frac{r^2}{l^2} - \frac{2 M}{r} \right)} dr^2 + r^2 d\Omega^2
\ee
The metric inside the collapsing shell is the usual in 3+1 dimensional AdS spacetime
\be
ds^2 = - \left(1  + \frac{r^2}{l^2}\right) dT^2 + {\left( 1 + \frac{r^2}{l^2} \right)}^{-1}dr^2 + r^2 d\Omega^2
\ee
The metric on the shell is given by
\be
ds^2 = - d \tau^2 + r^2 d\Omega^2
\ee
As in the section \ref{sc}, matching the time coordinates exactly at the shell gives us the relationship between the outside observer's time, $t$, and inside observer's time, $T$, with the proper time on the shell,$\tau$, as
\be
\frac{dT}{d\tau} = \frac{1}{B_{in}}\sqrt{ B_{in}+ \left(\frac{dR}{d\tau}\right)^2}
\ee
and
\be
\frac{dt}{d\tau} = \frac{1}{B_{out}}\sqrt{ B_{out}+ \left(\frac{dR}{d\tau}\right)^2}
\ee
where $B_{in} =( 1 + \frac{R^2}{l^2})$ and $B_{out} = (1 + \frac{R^2}{l^2} -  \frac{2M}{R})$.
Application of the Gauss-Codazzi method gives us the equation of motion of the collapsing shell as
\be
R_{\tau} = \sqrt{\left( \frac{2M + \frac{\mu^2}{R^2}}{2 \mu}\right)^2 - \frac{R^2}{l^2} -1}
\ee
Once the classical equation of motion is known, we can evaluate occupation number of the emitted radiation. Since the formalism and the setup are very similar to the 2+1 dimensional case, we directly show the results. In 3+1 dimensions, the temperature of the AdS black hole is given by
\be
T = \frac{1}{4 \pi r_H} + \frac{3 r_H}{4 \pi l^2} ,
\ee
where $r_H$ is the horizon radius of the black hole obtained by setting $g_{rr} = 0$. Small black holes with the horizon radius smaller than the AdS parameter $l$ behave like Schwarzschild black holes. They exhibit negative specific heat and are thermodynamically unstable. Black holes with the horizon radius larger than the AdS parameter $l$ behave exactly the opposite. Their temperature increases as mass grow, and they are thermodynamically stable. This transition between the small and large AdS black holes is known as the Hawking-Page phase transition\cite{Hawking:1982dh}. It is instructive to check if quantum radiation from the collapsing shell indicates the same behavior.

\begin{figure}[htpb]
\begin{center}
  \includegraphics[height=0.28\textwidth,angle=0]{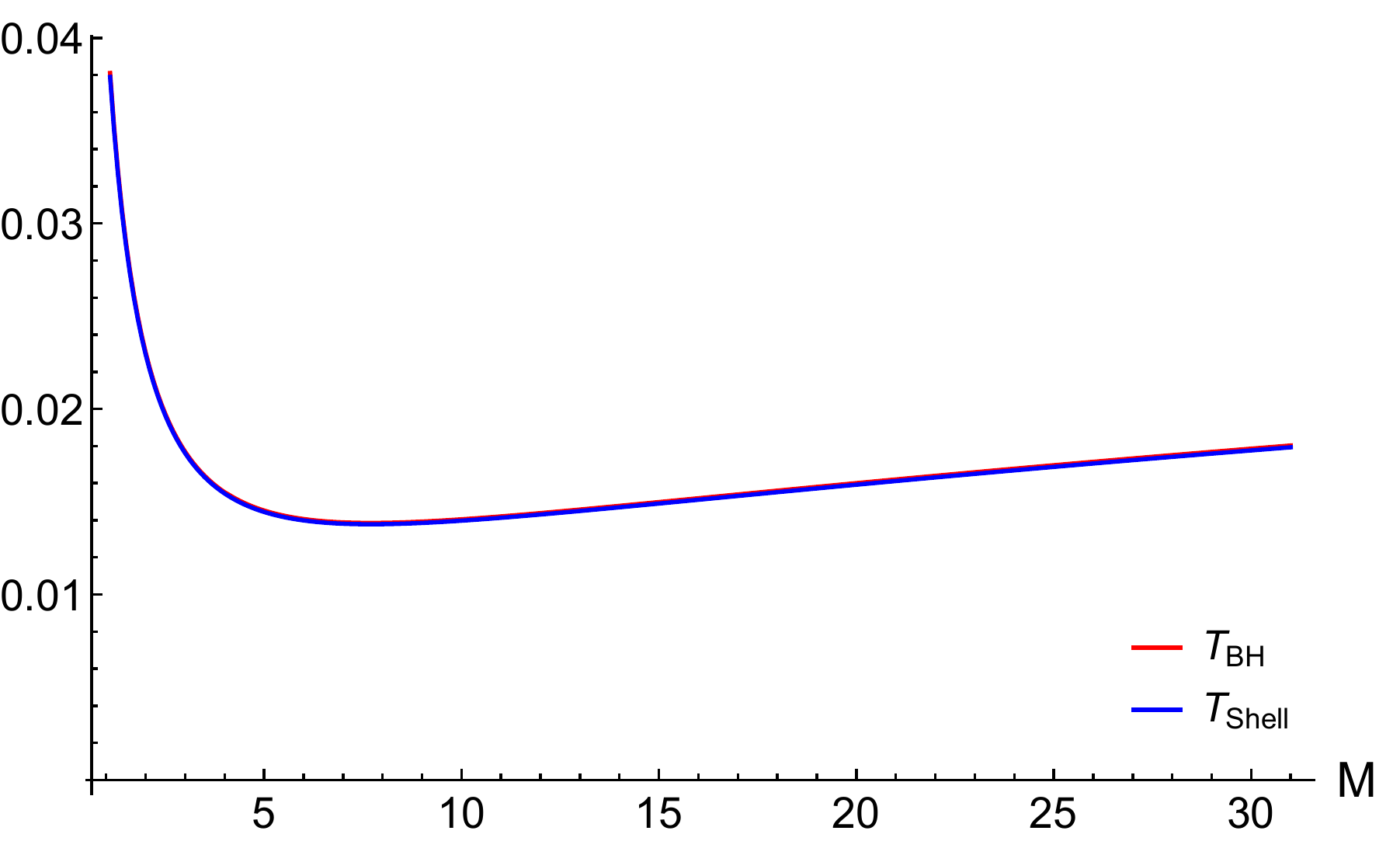}
\caption{We plot temperature of the radiation emitted by the collapsing shell in 3+1 dimensional AdS spacetime as a function of its mass, for the fixed value of the AdS parameter, $l = 20$. For small mass, the temperature is inversely proportional to the mass of the shell. However, for the massive shell, the temperature is directly proportional to the mass of the shell. This is in complete agreement with the standard Hawking-Page transition.}
\label{hawpage}
\end{center}
\end{figure}

In Fig.~\ref{hawpage}, we plot the temperature of the radiation from the collapsing shell at late times as a function of the mass of the shell, for the fixed value of the AdS parameter ($l =20$). When the corresponding horizon radius of the collapsing shell is very small, the temperature is inversely proportional to mass, indicating a negative specific heat. On the other hand, when the corresponding horizon radius of the collapsing shell is very large, the temperature is directly proportional to the mass of shell.

\begin{figure}[htbp]
\begin{center}
  \includegraphics[height=0.32\textwidth,angle=0]{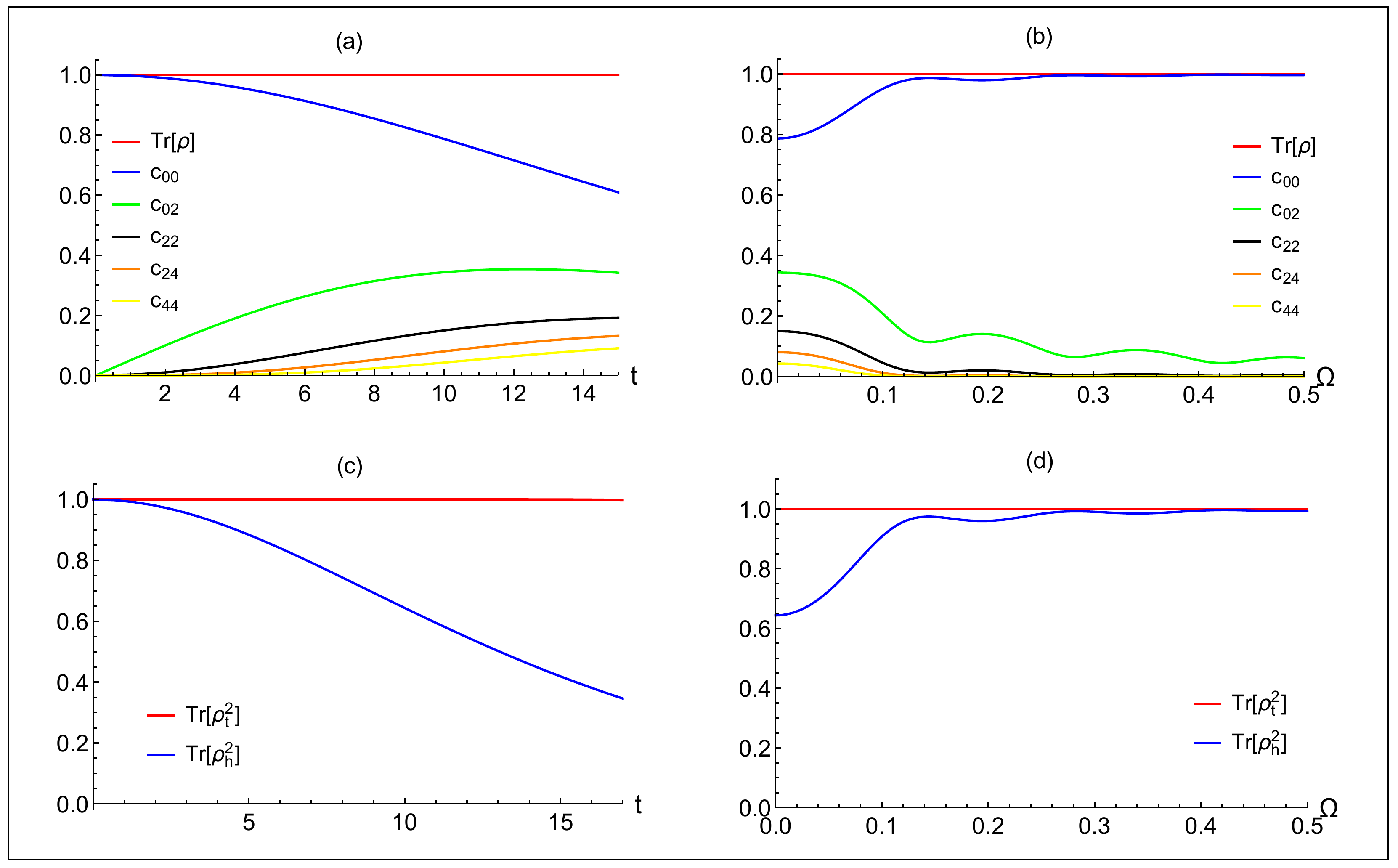}
\caption{Here, we plot all the relevant quantities for 3+1 dimensional asymptotic AdS spacetime. The results are very similar to the 2+1 dimensional case.}
\label{combined}
\end{center}
\end{figure}

In Fig.~\ref{combined},a, we plot various elements of the density matrix as a function of time at $\Omega = 0.001$. As we can see that the trend is similar to 2 + 1 dimensional case. Initially, the $c_{00}$ is unity and decreases with time. The contribution from the higher excited states increases as time progresses. In Fig.~\ref{combined},b, the elements of the density matrix are plotted as a function of frequency at $t =10$. The contribution from the vacuum, $c_{00}$, increases with the frequency which is reasonable since high-frequency modes are more difficult to excite. In Fig.~\ref{combined}, c and d, we plot Tr[$\rho^2$]  as a function of time and frequency. Like in 2+1 dimensional case, the off-diagonal elements are crucial in preserving unitarity of the system. The crucial difference is, however, that 3+1 dimensional AdS black holes are singular at the center, and the crucial question remains whether an outside observer can measure the complete density matrix even in principle.

\section{Conclusions}
\label{conclusions}
In this paper, we investigated a massive shell collapse in 2+1 and 3+1 asymptotically Anti-de Sitter spacetime. Using Gauss- Codazzi method, we studied classical dynamics of the shell in both space-times. The interaction between the time-dependent classical background and the quantum scalar field propagating in this background leads to the emission of particles. We employed the functional Schrodinger formalism to evaluate the wave function of the excited modes. This allowed us to study the time evolution of the system, which would not have been possible with standard Bogoliubov transformations. The calculated temperature of the emitted radiation at late times (when collapsing shell is very close to its horizon) was found to be in perfect agreement with the well-known Hawking's result, scaling as $T \propto \sqrt{M}/l$. The positive specific heat of the dust shell implies that in the 2+1 dimensional space-time, the dust shell can remain in equilibrium with the surrounding radiation. It is worth noting that the emitted radiation spectrum is completely Planckian only at low frequencies with respect to an outside observer.   At very high frequencies, the temperature diverges, possibly due to the non-renormalized action, and appropriate renormalization should fix this. At early times, quantum fluctuations are large, making it hard to draw any conclusions about the temperature of the emitted particle spectrum. 

We also constructed the density matrix of the excited modes in order to examine the information content of the emitted radiation. The correlations among excited modes are represented by the off-diagonal terms in the density matrix. We found that the off-diagonal elements of the density matrix are of the same order as the diagonal elements. The calculation of Tr[$\rho^2$] taking contributions only from the diagonal elements indicates the evolution of a pure state into a mixed state. However, Tr[$\rho^2$] of the density matrix containing all elements remains unity at all times and frequencies. This suggests that the off-diagonal terms are very important for the preservation of unitarity. Black holes in 2+1 dimensional AdS spacetime are special in the sense that they do not have a singularity at the center, so this result might imply that they do not suffer from the information loss problem. 

Finally, we considered 3+1 dimensional asymptotically AdS space-time. For a low mass shell, we found the temperature of the shell to be inversely proportional to its mass, whereas, for a heavier shell, the temperature was directly proportional to its mass. This perfectly corresponds to the Hawking-Page transition. We also evaluated the density matrix of the excited modes and found, similar to our results in 2+1 gravity spacetime, that off-diagonal terms cannot be neglected.

The findings in this paper indicate that the functional Schrodinger formalism fatefully reproduces results obtained by other methods, and at the same time has several advantages. It allows us to study the time evolution of the system and provides the complete wave function of the system with the total information about it.  

\begin{acknowledgments}
This work was partially supported by NSF, grant number PHY-1417317.
\end{acknowledgments}

\appendix

\end{document}